\documentclass[aps,floatfix,twocolumn,nofootinbib]{revtex4-1}
\usepackage{graphicx}
\usepackage[colorlinks = false,
            linkcolor = blue,
            urlcolor  = blue,
            citecolor = blue,
            anchorcolor = blue]{hyperref}

\begin{document}
\title{Boundary Orbits: 1 Static Spacetimes}
\author{Sheref Nasereldin and Kayll Lake }
\affiliation{Department of Physics, Queen's University, Kingston,
Ontario, Canada, K7L 3N6 }
\date{\today}

\begin{abstract}
The study of circular orbits in spacetime is of astrophysical importance. The identification and classification of circular orbits in both static and  stationary spacetimes remains an active area of interest. Even in the simplest static spherically symmetric case, it is well known that the introduction of a cosmological constant in vacuum leads to the study of quartic polynomials in order to locate \textit{boundary orbits}, those that straddle between stable and unstable orbits. These orbits are often referred to as ``marginally stable orbits'' or ``indifferently stable orbits". A comprehensive study of texts offers little clarification as to the stability or instability of these boundary orbits. Here we argue that the direct use of second order perturbation theory immediately shows that these boundary orbits are unstable in the perturbative sense. Our study here includes the two-particle Curzon-Chazy solution.
\end{abstract}

\maketitle
\section{Introduction}
The study of circular timelike geodesic orbits is a fundamental aspect of relativistic astrophysics. Determining the inner most stable circular orbits (ISCOs) in the fields of black holes and neutron stars provides information on the accretion disk region of these objects (e.g. \cite{Abram}), as the inner radius of the accretion disk is typically assumed to occur at the ISCO. During the accretion process the gas particles' gravitational potential energy is depleted and the gas is heated up. The amount of radiation energy in this process is equal to the gravitational binding energy of the ISCO \cite{Pringle}. The efficiency can be determined by dividing the gravitational binding energy of the ISCO by the rest mass energy of the particle \cite{Hobson}. ISCOs also play a crucial role in the field of gravitational waves as they mark the transition of the compact binaries from the slow inspiralling phase to the fast plunge phase \cite{Buon}. The gravitational waves which are generated by these sources can be classified according to the frequency at which the phase transition occurs  (see \cite{Damour1,Damour2} and below).

With the introduction of charge and/or a  cosmological constant, it is not difficult to see that outer most stable circular orbits (OSCOs) will also arise (e.g. \cite{stuchlik}). We refer to \textit{boundary orbits} as those that straddle between stable and unstable circular orbits and so include both ISCOs and OSCOs. (We avoid the use of the term ``marginally stable orbits''. See \cite{Beheshti}, \cite{Jia1} and \cite{Jia2}. The orbits are also referred to as ``indifferently stable". For example \cite{lukes}.)  With the existence of both ISCOs and OSCOs the search for boundary orbits leads directly to the study of polynomials. For example, recently Sturm's theorem has been used to study boundary orbits in the Kottler (Schwarzschild-de Sitter) spacetime \cite{Ono}.

Studying the existence and stability of circular orbits is, of course, a fundamental problem in dynamical systems and there are a number of approaches that can be used. Indeed, there are a number of types of stability \cite{Jackson}. For example, in addition to the use of the elementary ``effective potential" approach, it is now common to see the use of Lyapunov exponents (e. g. \cite{berti}). However, as we show, the direct use of second order perturbation theory immediately shows that boundary orbits are unstable in the perturbative sense and it is this type of instability we are interested in here. This instability is, at first sight, of no apparent deep physical significance\footnote{This is actually no longer true, given the observation of gravitational waves. ``Bumpy" spacetimes can be tuned arbitrarily close to the Kerr metric. However, orbits approaching an ISCO can be qualitatively different depending on whether the ISCO is determined by the onset of instability in the radial or vertical direction. See \cite{gair}. Here we are concerned with a direct treatment of the much simpler static case.}. However, it is certainly of pedagogical significance (think of the name, ISCO). Without a clear statement about the stability of boundary orbits, one has to skate awkwardly about the issue. For example, in one of the most influential texts of all time \cite{Gravitation}, the boundary orbits in Schwarzschild are described both as stable and unstable. In another influential text \cite{Chandrasekhar} the boundary orbits are described as stable in a figure but (and without justification) unstable in the text. A careful and interesting treatment is given in \cite{Woodhouse}, but the boundary orbits are not explicitly addressed. Generally speaking, texts consider only ISCOs, and they are considered ``stable". The same can be said of many, but not all, papers.

\section{The Static Spherically Symmetric Case}

\subsection{Background Calculations}

First we construct stable circular timelike geodesic orbits in a
static spherically symmetric field.  We follow the notation of \cite{lake}. The field takes the form
\begin{equation}
ds^2=\frac{dr^2}{1-\frac{2m(r)}{r}}+r^2d\Omega^2-e^{2\Phi(r)}dt^2,
\label{standardform}
\end{equation}
where $d\Omega^2$ is the metric of a unit sphere. Throughout we refer
to the function $\Phi(r)$ as the ``potential" and $m(r)$  as the effective gravitational ``mass". It is immediately clear from
(\ref{standardform}) that all geodesic orbits are stably planar
(say $\theta=\pi/2$) and have two constants of motion, the
``energy" $\gamma=e^{2\Phi(r)}\dot{t}$ ( $\dot{}$ being the proper time derivative) and ``angular momentum"
$l=r^2\dot{\phi}$. In the timelike case then
\begin{equation}
\dot{r}^2f(r)+\mathcal{V}(r)=\gamma^2, \label{effective}
\end{equation}
where
\begin{equation}
f(r)=\frac{e^{2\Phi(r)}}{1-\frac{2m(r)}{r}} > 0\label{function}
\end{equation}
and
\begin{equation}
\mathcal{V}(r)=e^{2\Phi(r)}(1+\frac{l^2}{r^2}).\label{effectivepotential}
\end{equation}

Setting $\dot{r}=\ddot{r}=0,\; r>0$ it follows from the timelike
geodesic equations that
\begin{equation}
\gamma^2=\frac{e^{2 \Phi}}{1-r\Phi^{'}}\label{energy}
\end{equation}
and
\begin{equation}
l^2=\frac{r^3\Phi^{'}}{1-r\Phi^{'}}\label{angularmomentum}
\end{equation}
where $^{'} \equiv d/dr$. The existence of these circular
orbits requires
\begin{equation}
0<r\Phi^{'}<1.\label{orbits}
\end{equation}
Note that from (\ref{effectivepotential}) and (\ref{angularmomentum}) on a circular orbit ($r=r_{0}$)
\begin{equation}\label{vprime}
    \mathcal{V}^{'}(r_{0})=0
\end{equation}
and
\begin{equation}\label{vdprime}
    \mathcal{V}^{''}(r_{0})=2e^{2\Phi}\left(\frac{3\Phi^{'}+r\Phi^{''}-2r(\Phi^{'})^{2}}{r(1-r\Phi^{'})}\right).
\end{equation}

\subsection{Radial Perturbations}

Next, we require that the timelike circular geodesics be stable against radial perturbations. Other perturbations (in energy and angular momentum) are of no interest here as explained below.
Let $r_{0}$ be a circular orbit and consider $r=r_{0}+\delta$
where $\delta<<r_{0}$. Taking expansions\footnote{The expansions do not involve reference to the metric components and so (\ref{stable}) and (\ref{marginal}) apply to the axial case discussed below.} of $\mathcal{V}(r)$ and
$f(r)$ about $r=r_{0}$ it follows from (\ref{effective}) that to order $\delta$
\begin{equation}
\ddot{\delta}+\frac{\mathcal{V}^{''}(r_{0})}{2f(r_{0})}\delta=0\label{stable},
\end{equation}
so that $\mathcal{V}^{''}(r_{0})>0$ for stability. From (\ref{vdprime}) then
\begin{equation}
3\Phi^{'}+r\Phi^{''}>2r(\Phi^{'})^{2}\label{stability}
\end{equation}
for stable circular orbits. Clearly $\mathcal{V}^{''}(r_{0})<0$ leads to instability. The case $\mathcal{V}^{''}(r_{0})=0$ defines the boundary orbits. To examine stability in this case we must go to order $\delta^2$. We then find
\begin{equation}
\ddot{\delta}+\frac{\mathcal{V}^{'''}(r_{0})}{2f(r_{0})}\delta^2=0,\label{marginal}
\end{equation}
and so the solutions for $\delta$ are Weierstrass {\large$\wp$} functions (e.g. \cite{Abramowitz}). We therefore interpret these orbits as unstable and we refer to the boundary orbits as $r_{*}$ where $\mathcal{V}^{''}(r_{*})=0$.

\subsection{Newtonian Comparison and a glance at More Involved Boundary Orbits}

The stability condition (\ref{stability}) can be considered a refinement of the Newtonian
condition
\begin{equation}
3\Phi^{'}+r\Phi^{''}>0\label{newtonian}
\end{equation}
for conservative central fields. In Table 1 we compare conditions (\ref{stable}) and (\ref{newtonian}) in the simplest cases, a point mass in Newton's theory and the Schwarzschild solution\footnote{ Stricktly speaking, of course, only (\ref{newtonian}) is appropriate in Newtonian theory and (\ref{stability}) in general relativity. However, we include this table to show, for example, that Newton's theory somehow ``knows" that with the correct stability criterion, $r>2m$ for stable circular orbits.}
\begin{table}[ht]
\caption{Comparison}
\centering
\begin{tabular}{c c c c}
\hline\hline
Theory & $\Phi$ & (\ref{stability}) & (\ref{newtonian}) \\ [0.5ex]
\hline
Newton&$-m/r$&$r>2 m$&$m > 0$ \\
Einstein&$\frac{1}{2}\ln(1-\frac{2 m}{r})$&$r > 6 m$&$ r > 4 m$ \\ [1ex]
\hline
\end{tabular}
\label{table:nonlin}
\end{table}

\bigskip

These simple results quickly give way to more involved relations. For example, an inclusion of the cosmological constant (here we take $\Lambda \geq 0$) gives rise to the Newtonian potential $\Phi(r)=-m/r-\Lambda r^2/6$ and the Kottler (Schwarschild - de Sitter) solution $\Phi(r)=\frac{1}{2}\ln(1-2 m/r - \Lambda r^2/3)$. In the Newtonian case the boundary orbits satisfy $3\Phi^{'}(r_{*})+r\Phi^{''}(r_{*})=0$ and so
\begin{equation}\label{newton2}
r_{*} = \sqrt[3]{\frac{3 m}{4 \Lambda}}.
\end{equation}
The Newtonian circular orbits are stable for $r < r_{*}$, with $r_{*}$ given by (\ref{newton2}).

In the Kottler case the boundary orbits $\mathcal{V}^{''}(r_{*})=0$ give
\begin{equation}\label{kottler}
4 \Lambda r_{*}^4-15 m \Lambda r_{*}^3-3m r_{*}+18 m^2 = 0,
\end{equation}
so that $r_{*}$ has up to 2 distinct values. As mentioned previously, equation (\ref{kottler}) has been examined recently via Sturm's theorem by Ono \textit{et al.} \cite{Ono}. The inclusion of charge (the Reissner Nordstr\"{o}m de Sitter solution with $\Phi(r)=\frac{1}{2}\ln(1-2 m/r - \Lambda r^2/3 + e^2/r^2)$) requires numerical methods to determine the boundary orbits. These are given by
\begin{eqnarray}\label{rnds}
4 \Lambda r_{*}^6-15 m \Lambda r_{*}^5+12 \Lambda e^2 r_{*}^4-3m r_{*}^3+\nonumber \\ 18m^2r_{*}^2-27me^2r_{*}+12e^4 = 0
\end{eqnarray}
so that $r_{*}$ has up to 6 distinct values.

\subsection{Changes in Energy and Angular Momentum}

Given $r_{*}$, $\Phi^{'}_{*}$ and $\Phi_{*}$, $\gamma_{*}$ and $l_{*}$ are given uniquely by (\ref{energy}) and (\ref{angularmomentum}) respectively. The energy and angular momentum of boundary orbits is therefore predetermined. Note also that $\gamma^{'}_{*}=l^{'}_{*}=0$ but the nature of the associated critical points depends on $\Phi^{'''}_{*}$. The point here is that when considering perturbations about the boundary orbits $r_{*}$, changes in $\gamma$ and $l$ can not be considered.

\bigskip

\section{The Static Axially Symmetric Case}

\subsection{A Metric of Sufficient Generality}

We start with the axially symmetric metric
\begin{equation}\label{axial}
  ds^2 = 2 H dt d\phi + B^2 dr^2 + C^2 d\theta^2 + F^2 d\phi^2-e^{2 \Phi}dt^2
\end{equation}
where $H$, $B$, $C$, $F$ and $\Phi$ are functions of $r$ and $\theta$. The Killing vector $\xi^{\alpha} = \delta^{\alpha}_{t}$ satisfies
\begin{equation}\label{static}
  \xi_{[ \alpha}\nabla_{\beta}\xi_{\delta ]} = 0
\end{equation}
if and only if $H=0$ which is our static condition.

\bigskip

\subsection{Equatorial Orbits in Reflection Symmetric Spacetimes}

By equatorial orbits we mean orbits in the subspace $\theta=\pi/2$ and by reflection symmetric we mean
\begin{equation}\label{reflection}
  \frac{\partial X}{\partial \theta}\bigg|_{\theta=\pi/2} = 0
\end{equation}
where $X \in (B, C, F, \Phi)$. Under these conditions, and at a superficial level, there is relatively little change from the spherical case.
Whereas (\ref{effective}) remains unchanged, equations (\ref{function}) through (\ref{angularmomentum}) become
\begin{equation}
f=B^2 e^{2\Phi(r)} > 0,\label{function'}
\end{equation}
\begin{equation}
\mathcal{V}=e^{2\Phi}(1+\frac{l^2}{F^2}),\label{effectivepotential'}
\end{equation}
\begin{equation}
\gamma^2=\frac{e^{2 \Phi} F^{'}}{F^{'}-F\Phi^{'}}\label{newenergy}
\end{equation}
and
\begin{equation}
l^2=\frac{F^3\Phi^{'}}{F^{'}-F\Phi^{'}}\label{newangularmomentum}
\end{equation}
respectively where $^{'} \equiv \partial /\partial r |_{\theta=\pi/2}$. Moreover, whereas (\ref{vprime}) remains unchanged, (\ref{vdprime}) becomes
\begin{equation}\label{vdprime'}
\mathcal{V}^{''}=2e^{2\Phi}\left(\frac{(3(F^{'})^2-F F{''})\Phi^{'}+F F^{'}(\Phi^{''}-2(\Phi^{'})^{2})}{F(F^{'}-F\Phi^{'})}\right).
\end{equation}
The condition for the stability of circular orbits remains $\mathcal{V}^{''}(r_{0})>0$ and boundary orbits are once again given by $\mathcal{V}^{''}(r_{*})=0$.

\subsection{Vacuum}

The static axially symmetric vacuum solutions of the Einstein equations are the Weyl metrics\footnote{Weyl metrics do not allow for $\Lambda$ as $G_{\rho\rho} = -G_{zz}$ which necessitates $\Lambda = 0$.}\cite{weyl}
\begin{equation}\label{weyl}
ds^2= -e^{2U}dt^2+e^{-2U}(e^{2\gamma}(d\rho^2+dz^2)+\rho^2d\phi^2),
\end{equation}
where $U$ and $\gamma$ are functions of $\rho$ and $z$. The simplest case is the Zipoy \cite{zipoy} - Voorhees \cite{voorhees} solution which has been very widely studied (see, for example, \cite{lukes}).

The next most widely studied case is the Curzon \cite{curzon} - Chazy \cite{chazy} solution. (For a recent study of the geometrical properties of this solution see \cite{majd}. For a recent study of timelike geodesic orbits in this solution see \cite{dolan}.) In Stachel coordinates \cite{stachel} the solution is given by
\begin{eqnarray}\label{stachel}
ds^2=e^{\frac{2 m}{r}}(e^{-\frac{m^2 \sin(\theta)^2}{r^2}}(d r^2 +r^2 d \theta^2)\nonumber \\
+r^2 \sin(\theta)^2 d\phi^2)-e^{-\frac{2 m}{r}} d t^2,
\end{eqnarray}
where $\rho=r \sin(\theta)$ and $z = r \cos(\theta)$.
From (\ref{vdprime'}) and (\ref{stachel}) it now follows that
\begin{equation}\label{stachel1}
\mathcal{V}^{''} = \frac{2(x^2-6x+4)}{x^4e^{2/x}(x-2)}
\end{equation}
where $x \equiv r/m$. Even though the metric (\ref{stachel}) remains regular at $r=2m$ \cite{majd}, the spacetime is divided into two regions as regards circular timelike orbits in the equatorial plane. For $r>2m$, $x_{*} = 3+\sqrt{5}$ and the timelike circular orbits are stable for $x>x_{*}$ and unstable for $x \leq x_{*}$. The minimum value of $l^2$ is at $l_{*}^2$ which is $>0$. For $r=2m$ neither $\gamma$ nor $l$ exist. For $r<2m$, $l$ does not exist. There is then only one boundary orbit \cite{deFelice}, $r_{*}=(3+\sqrt{5})m\cong5.24m$. This is all remarkably similar to the Schwarzschild vacuum, with $2m$ replaced by $3m$ and $(3+\sqrt{5})m$ replaced by $6m$. Since the boundary orbits in the one - particle Curzon - Chazy solution is so simple, one is encouraged to look at the two - particle case \cite{griffiths}.

In the two - particle case we have
\begin{equation}
U=-\frac{m1}{\sqrt{\rho^2+(z-a)^2}}-\frac{m2}{\sqrt{\rho^2+(z+a)^2}}
\end{equation}
and
\begin{equation}
F=\rho e^{-U}.
\end{equation}

Once again, using $\rho=r \sin(\theta)$ and $z = r \cos(\theta)$, with $A = a/(m1+m2)$ and $k = \sqrt{r^2+a^2}/(m1+m2)$, it follows that in the equatorial plane $\mathcal{V}^{''}(m1+m2)^2$ is given by
\begin{equation}\label{twop}
 \frac{2(k^6-6k^5+k^4(3A^2+4)+6A^2k^3-8A^2k^2+4A^4)}{e^{2/k}(k^3+2(A-k)(A+k))k^6}.
\end{equation}
The single particle case is given by $A=0$ and $m = m1+m2$. Note that $k \geq A$ and
\begin{equation}
\mathcal{V}^{''}(m1+m2)^2 \bigg|_{k=A} = \frac{8}{A^3e^{2/A}}.
\end{equation}

From (\ref{newenergy}) and (\ref{newangularmomentum}) we now find that
\begin{equation}\label{2newenergy}
\gamma^2 = \frac{k^3+(A-k)(A+k)e^{-2/k}}{k^3+2(A-k)(A+k)}
\end{equation}
and
\begin{equation}\label{2newangularmomentum}
\frac{l^2}{(m1+m2)^2} = \frac{(A-k)^2(A+k)^2e^{2/k}}{k^3+2(A-k)(A+k)}
\end{equation}

Clearly the polynomial
\begin{equation}\label{poly}
k^3+2(A-k)(A+k),
\end{equation}
plays an important role in (\ref{twop}), (\ref{2newenergy}) and (\ref{2newangularmomentum}). The polynomial distinguishes two ranges in $A$. For $A > 4/3\sqrt{3}$, (\ref{poly}) has no positive roots. For $A = 4/3\sqrt{3}$ there is one positive root given by $k=4/3$. For $0 < A < 4/3\sqrt{3}$ there are two positive roots. Call them $k_{1}$ and $k_{2}$. Between these roots $k^3+2(A-k)(A+k)<0$ and so $l$ does not exist. The polynomial (\ref{poly}) is discussed in the Appendix.

\bigskip

Properties of (\ref{twop}) can be summarized as follows:
\begin{itemize}
\setlength\itemsep{-0.5em}
\item For $A>A_{0}\simeq 1.095886$, boundary orbits $r_{*}$ do not exist. All circular orbits are stable.

\item For $A=A_{0}$, one boundary orbit exists with $k_{*}\simeq 3.092671$. Otherwise, all circular orbits are stable.

\item For $4/3\sqrt{3} < A < A_{0}$ there exist two boundary orbits with, say, $k_{*} = k_{*1}$ and $k_{*} = k_{*2}$ where $k_{*1} < k_{*2}$. The circular orbits are stable for $k < k_{*1}$ and $k > k_{*2}$.

\item For $4/3\sqrt{3} = A$, the circular orbits are stable for $k < 4/3\sqrt{3}=k_{*1}$ and $k> k_{*2} \simeq 4.634848$.

\item For $0 < A < 4/3\sqrt{3}$ recall that the denominator has two roots, $k_{1}$ and $k_{2}$. Circular orbits are stable for $0<k<k_{1}$ and $k_{2} < k_{*2} < k$.
\end{itemize}

\textit{Acknowledgments.} This work was supported by a grant (to KL) from the Natural Sciences and Engineering Research Council of Canada.

\bigskip

\appendix*

\section{Properties of (\ref{poly})}
The property of (\ref{poly}) that no positive roots exist for $A > 4/3\sqrt{3}$ follows immediately from the following elementary theorem \cite{levin}: For
\begin{equation}\label{levin}
    1-ax^m+bx^n
\end{equation}
$a, b$ real and $n>m>0$ integers, there are no positive roots as long as $a>0$ and
\begin{equation}\label{levin1}
    b>(\frac{a}{\alpha})^{\alpha}(\alpha-1)^{\alpha-1}
\end{equation}
where $\alpha \equiv n/m$.


\begin{thebibliography}{}\label{sec:TeXbooks}
\bibitem{Abram} M. A. Abramowicz, M. Jaroszy\'{n}ski, S. Kato, J.-P. Lasota, A. R\'{o}\.{z}a\'{n}ska, and A. S\c{a}dowski, Astronomy and Astrophysics 521, (2010) A15.
\bibitem{Pringle} J. E. Pringle, Annual Review of Astronomy and Astrophysics \textbf{19}, 137 (1981).
\bibitem{Hobson} M. P. Hobson, G. P. Efstathiou, and A. N. Lasenby, \textit{General Relativity an Introduction for Physicists} (Cambridge University Press, Cambridge, UK, 2014).
\bibitem{Buon}A. Buonanno, L. E. Kidder, and L. Lehner, Phys. Rev. D \textbf{77},026004 (2008),[arXiv: astro-ph/0709.3839].
\bibitem {Damour1} T. Damour, B. R. Iyer, and B. S. Sathyaprakash, Physical Review D \textbf{57}, 885 (1998).
\bibitem{Damour2}T. Damour, P. Jaranowski, and G. Schäfer, Physical Review D \textbf{62}, 084011 (2000).
\bibitem{stuchlik} Z. Stuchlik and S. Hledik, Phys. Rev. D, \textbf{60} 044006 (1999).
\bibitem{Beheshti} S. Beheshti and E. Gasperín, Physical Review D \textbf{94}, 024015 (2016).
\bibitem{Jia1} J. Jia, J. Liu, X. Liu, Z. Mo, X. Pang, Y. Wang, and N. Yang, General Relativity and Gravitation \textbf{50}, 40 (2018).
\bibitem{Jia2} J. Jia, X. Pang, and N. Yang, General Relativity and Gravitation \textbf{50}, 17 (2018).
\bibitem{lukes} G. Lukes-Gerakopoulos, Phys. Rev. D, \textbf{86} 044013 (2012), [arXiv: gr-qc/1206.0660].
\bibitem{Ono} T. Ono, T. Suzuki, N. Fushimi, K. Yamada and H. Asada, Europhys. Lett \textbf{111}, 3008 (2015), [arXiv:gr-qc/1508.00101].
\bibitem{Jackson} E. Atlee Jackson, \textit{Perspectives of nonlinear dynamics, 1} (Cambridge University Press, Cambridge, UK, 1990).
\bibitem{berti} ``\textit{A Black-Hole Primer: Particles, Waves, Critical Phenomena and Superradiant Instabilities}", Bad Honnef School "GR@99", [arXiv: gr-qc/1410.4481]. For the Schwarzschild vacuum, Lyapunov stability is known to be equivalent to Jacobi stability. See H. Abolghasem, International Journal of Differential Equations and Applications, \textbf{12}, 131 (2013).
\bibitem{gair} J. R. Gair, C. Li, and I. Mandel, Phys. Rev. D, \textbf{77}, 024035 (2008), [arXiv: gr-qc/0708.0628].
\bibitem{Gravitation} C. W. Misner, K. S. Thorne and J. A. Wheeler, \textit{Gravitation} (Princeton University Press, Princeton, NJ, 2017). This is a reprint with pagination identical to the original Freeman version. The boundary orbits in Schwarzschild are described both as stable (section 6) and unstable (section 7) in Box 25.6 page 662.
\bibitem{Chandrasekhar} S. Chandrasekhar,\textit{ The Mathematical Theory of Black Holes} (Clarendon Press, Oxford, UK, 2009). See figure 5 and equation (144') in chapter 3.
\bibitem{Woodhouse} N. M. J. Woodhouse, \textit{General Relativity} (Springer-Verlag London Limited, New York, 2007).
\bibitem{lake} K. Lake, Physical Review Letters \textbf{92}, 051101 (2004), [arXiv:gr-qc/0302067].
\bibitem{Abramowitz} M. Abramowitz and I. A. Stegun, \textit{Handbook of mathematical functions}. Many versions of this classic work are available on the internet.
\bibitem{weyl}H. Weyl, Ann. Phys. (Berlin) \textbf{364}, 185 (1919).
\bibitem{zipoy} D. M. Zipoy, Journal of Mathematical Physics \textbf{7}, 1137 (1966).
\bibitem{voorhees} B. H. Voorhees, Phys. Rev. D2, 2119 (1970).
\bibitem{curzon} H. E. J. Curzon, Proc. London Math. Soc. \textbf{s2-23}, 477 (1925).
\bibitem{chazy} J. Chazy, Bull. Soc. Math. Fr. \textbf{52}, 17 (1924).
\bibitem{majd} M. Abdelqader and K. Lake, Phys. Rev. D 86, 124037 (2012), [arXiv:gr-qc/1207.5496].
\bibitem{dolan} S. R. Dolan, \textit{``Comment on Geodesic dynamics on Chazy-Curzon spacetimes"}, [arXiv:gr-qc/1901.01202].
\bibitem{stachel} J. Stachel, Phys. Lett. \textbf{27A}, 60 (1968).
\bibitem{deFelice} F. de Felice, General Relativity and Gravitation \textbf{23}, 135 (1991).
\bibitem{griffiths} J. B. Griffiths and J. Podolsk\'{y}, \textit{Exact Space-Times in Einstain's General Relativity} (Cambridge University Press, Cambridge, UK, 2009)
\bibitem{levin} A proof by S. A. Levin can be found at \texttt{http://sepwww.stanford.edu/oldsep/stew/descartes.pdf}
\end{thebibliography}
\end{document}